\documentclass[conference]{IEEEtran}
\IEEEoverridecommandlockouts
\usepackage{amsmath,amsfonts,amssymb, amsthm}
\usepackage{subcaption}
\usepackage[ruled,vlined,linesnumbered]{algorithm2e}
\usepackage{array}
\usepackage{textcomp}
\usepackage{stfloats}
\usepackage{url}
\usepackage{verbatim}
\usepackage{graphicx}
\usepackage{float}
\usepackage{cite}
\usepackage{graphicx}
\usepackage{xcolor}
\usepackage{color}
\usepackage[caption=false,font=footnotesize,labelfont=rm,textfont=rm]{subfig}
\usepackage[font=footnotesize,labelfont=rm,textfont=rm]{caption}
\def\BibTeX{{\rm B\kern-.05em{\sc i\kern-.025em b}\kern-.08em
		T\kern-.1667em\lower.7ex\hbox{E}\kern-.125emX}}
\begin{document}
	
	\title{Joint Association and Phase Shifts Design for UAV-mounted Stacked Intelligent Metasurfaces-assisted Communications\vspace{-12pt}}
	
	\author{\IEEEauthorblockN{
			Mingzhe Fan\IEEEauthorrefmark{1},
			Geng Sun\IEEEauthorrefmark{1}\IEEEauthorrefmark{2},
			Hongyang Pan\IEEEauthorrefmark{3},
			Jiacheng Wang\IEEEauthorrefmark{2},
			Jiancheng An\IEEEauthorrefmark{4},
			Hongyang Du\IEEEauthorrefmark{5},
			Chau Yuen\IEEEauthorrefmark{4}
		}
		
		\IEEEauthorblockA{\IEEEauthorrefmark{1}College of Computer Science and Technology, Jilin University, Changchun, China}
		\IEEEauthorblockA{\IEEEauthorrefmark{2}School of Computer Science and Engineering, Nanyang Technological University, Singapore, Singapore}
		\IEEEauthorblockA{\IEEEauthorrefmark{3}Information Science and Technology College, Dalian Maritime University, Dalian, China}
		\IEEEauthorblockA{\IEEEauthorrefmark{4}School of Electrical and Electronics Engineering, Nanyang Technological University, Singapore, Singapore}
		\IEEEauthorblockA{\IEEEauthorrefmark{5}Department of Electrical and Electronic Engineering, University of Hong Kong, Hong Kong, China}
		Email: sungeng@jlu.edu.cn,  panhongyang18@foxmail.com\\
		\IEEEauthorrefmark{2}Corresponding author: Hongyang Pan
		\vspace{-20pt}}
	
	\markboth{Journal of \LaTeX\ Class Files,~Vol.~14, No.~8, August~2021}%
	{Shell \MakeLowercase{\textit{et al.}}: A Sample Article Using IEEEtran.cls for IEEE Journals}
	
	\maketitle
	\addtolength{\topmargin}{0.01in}
	
	\begin{abstract}
		Stacked intelligent metasurfaces (SIMs) have emerged as a promising technology for realizing wave-domain signal processing, while the fixed SIMs will limit the communication performance of the system compared to the mobile SIMs. In this work, we consider a UAV-mounted SIMs (UAV-SIMs) assisted communication system, where UAVs as base stations (BSs) can cache the data processed by SIMs, and also as mobile vehicles flexibly deploy SIMs to enhance the communication performance. To this end, we formulate a UAV-SIM-based joint optimization problem (USBJOP) to comprehensively consider the association between UAV-SIMs and users, the locations of UAV-SIMs, and the phase shifts of UAV-SIMs, aiming to maximize the network capacity. Due to the non-convexity and NP-hardness of USBJOP, we decompose it into three sub-optimization problems, which are the association between UAV-SIMs and users optimization problem (AUUOP), the UAV location optimization problem (ULOP), and the UAV-SIM phase shifts optimization problem (USPSOP). Then, these three sub-optimization problems are solved by an alternating optimization (AO) strategy. Specifically, AUUOP and ULOP are transformed to a convex form and then solved by the CVX tool, while we employ a layer-by-layer iterative optimization method for USPSOP. Simulation results verify the effectiveness of the proposed strategy under different simulation setups.
	\end{abstract}
	
	\vspace{-0.2cm}
	
	%
	%
	\section{Introduction}
	\vspace{-0.1cm}
	
	\par Stacked intelligent metasurfaces (SIMs) have recently gained attention as an innovative technology, which can be used in edge computing systems and internet-of-things networks \cite{DBLP:journals/jsac/AnXNAHYH23}. Specifically, SIM can be structured similarly to artificial neural networks by stacking multiple programmable metasurfaces, thereby obtaining a better signal processing capability compared to the single-layer structure \cite{DBLP:journals/wcl/PapazafeiropoulosKCKV24}. Moreover, SIM performs forward propagation at the speed of light without incurring the extra processing delay. Therefore, it can accomplish advanced computation and signal processing tasks by manipulating the electromagnetic (EM) waves. Besides, SIM can also reduce the radio frequency chain while maintaining the same beamforming gain, thus eliminating the reliance on high-precision analog-to-digital and digital-to-analog conversion components \cite{DBLP:conf/icc/AnRDY23}. 
	
	\par Since SIM systems usually have no cache modules, it is generally better to integrate a separate base station (BS) for each SIM \cite{DBLP:journals/jsac/AnXNAHYH23}. Besides, SIM enhances BS performance by enabling wave-based beamforming. However, deploying SIM on a fixed BS presents significant limitations, such as limited coverage and lack of flexibility. Thus, it is essential to configure a mobile BS for each SIM to further enhance the communication performance of the system. 
	
	\par Among the mobile BSs, unmanned aerial vehicles (UAVs) have garnered significant interest due to their high flexibility, low-cost and fast deployment \cite{pan2023joint}. Moreover, UAVs can also introduce the high-quality line-of-sight channel conditions \cite{Zhang2025}, which is helpful for the signal processing of SIMs. Furthermore, UAVs have a certain cache capacity, which can store the data obtained through SIM-based wave-domain calculation. Due to the mutual benefits, UAV-mounted SIMs (UAV-SIMs) will be a promising technology in 5G/6G networks\footnote{SIM is generally structured with multiple printed circuit boards, and hence it is lightweight and can be conveniently mounted on UAVs \cite{DBLP:journals/corr/abs-2405-01104}.}, where UAVs act as the mobile BSs and SIMs serve as transceivers, which can ultimately enhance communication performance.
	
	\par This paper investigates a UAV-SIMs-assisted uplink communication system, and the main contributions are summarized as follows: 
	\begin{itemize}
		\item We consider an uplink communication scenario where multiple UAV-SIMs are deployed to provide link service for the ground users. In the considered system, we formulate a UAV-SIM-based joint optimization problem (USBJOP) to comprehensively consider the association between UAV-SIMs and users, the locations of {\spaceskip=0.2em\relax UAV-SIMs, and the phase shifts of UAV-SIMs, aiming to} maximize the network capacity, which is non-convex and NP-hard. 
		\item Due to the complexity of USBJOP, it is decomposed into three sub-optimization problems, which are the association between UAV-SIMs and users optimization problem (AUUOP), the UAV location optimization problem (ULOP), and the UAV-SIM phase shifts optimization problem (USPSOP). Then, an alternating optimization (AO) strategy is proposed to solve them iteratively. 
		\item We perform simulation experiments, demonstrating that UAV-SIMs relying on the proposed AO strategy effectively enhance network capacity. Specifically, the proposed AO strategy performs approximately two times better than the suboptimal benchmark.
	\end{itemize}
	
	\par The rest of this paper is organized as follows. Section \ref{sec2} introduces the system model and formulates the USBJOP, and then an AO strategy is proposed in Section \ref{section3} to solve the USBJOP. Furthermore, the simulation results are presented in Section \ref{sec4}. Finally, the conclusions of the paper are presented in Section \ref{sec5}.
	
	\par \textit{Notations:} In this paper, $(\cdot)^{\mathrm{H}}$ and $(\cdot)^{\mathrm{T}}$ denote the Hermitian transpose and transpose, respectively; bold uppercase and lowercase letters denote matrices and vectors, respectively; $ \operatorname{diag}(\cdot)$ means a diagonal matrix; the smallest integer greater than or equal to $x$ is denoted as $\lceil x\rceil$; the space of $x \times y$ complex-valued matrices is represented by $\mathbb{C}^{x \times y}$; $\|\mathbf{x}\|$ represents the norm of the vector $x$; mod $(x, n)$ returns the remainder after division of $x$ by $n$; the distribution of a circularly symmetric complex Gaussian random vector with mean vector $\boldsymbol{\mu}$ and covariance matrix $\boldsymbol{\Xi} \succeq \mathbf{0}$ is denoted by $\sim \mathcal{C N}(\boldsymbol{\mu}, \boldsymbol{\Xi})$, where $\sim$ stands for ``distributed as''; $\operatorname{sinc}(x)=\frac{\sin (\pi x)}{\pi x}$ is the normalized sinc function; $j$ is the imaginary unit satisfying $j^2=-1$.
	\vspace{-2pt}
	
	%
	%
	\section{System Model and Problem Formulation}\label{sec2}
	
	\subsection{Network Model}
	\begin{figure*}[tb]
		\centering
		\begin{subfigure}{0.495\linewidth}
			\centering
			\includegraphics[width=0.92\linewidth]{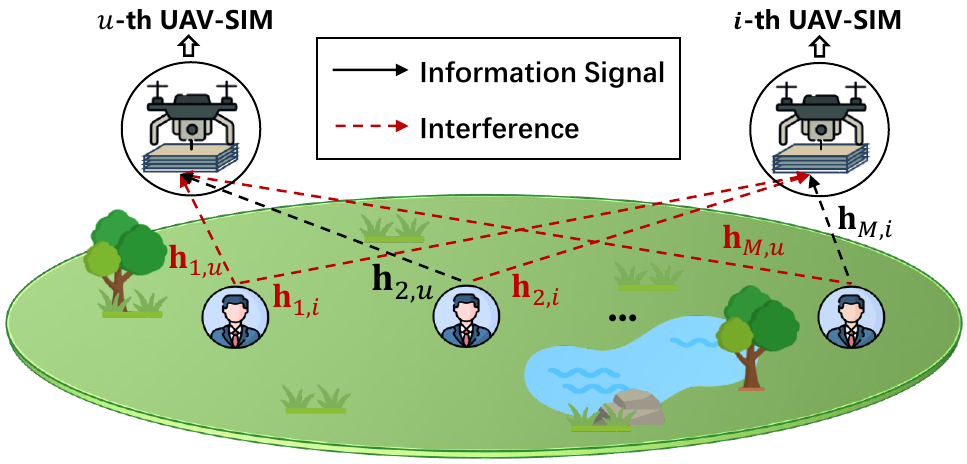}
			\caption{}
			\label{A UAV-mounted SIM architecture}
		\end{subfigure}
		\centering
		\begin{subfigure}{0.495\linewidth}
			\centering
			\includegraphics[width=0.92\linewidth]{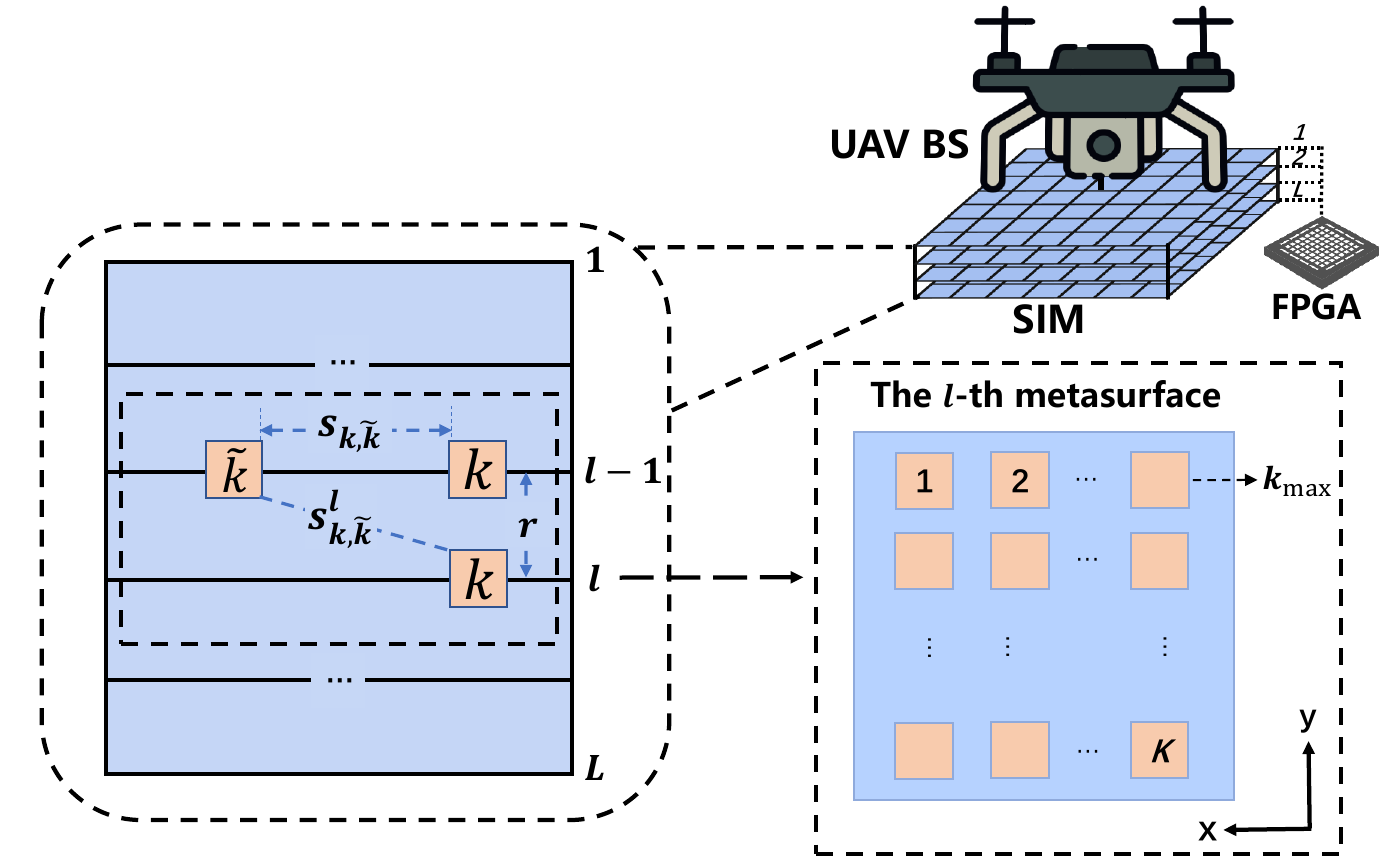}
			\caption{}
			\label{UAV-mounted SIMs assisted uplink communications}
		\end{subfigure}
		\caption{(a) UAV-SIMs-assisted uplink communications. (b) Architecture of a UAV-SIM.}
		\label{System_model}
		\vspace{-15pt}
	\end{figure*}
	
	\par As depicted in Fig. \ref{System_model}(a), we consider UAV-SIMs-assisted uplink communications and the architecture of a UAV-SIM is shown in Fig. \ref{System_model}(b). Specifically, a set of UAV-SIMs $\mathcal{U} = \{1, \ldots, U\}$ are deployed to serve a group of ground users $\mathcal{M} = \{1, \ldots, M\}$. Without loss of generality, we assume that $U<M$. Moreover, each UAV-SIM is integrated with a single antenna. Since all UAV-SIMs share the same frequency, mutual interference occurs in the considered communication system. 
	
	\par We assume that the $m$-th user is fixed at $\mathbf{p}_m=[x_m,y_m,0]^{\mathrm{T}}, m \in \mathcal{M}$, and all UAV-SIMs fly at a fixed altitude $H$. Then, the three-dimensional (3D) coordinate of the $u$-th UAV-SIM can be represented as $\mathbf{q}_u=[x_u,y_u,H]^{\mathrm{T}}, u \in \mathcal{U}$. Additionally, UAV-SIMs need to maintain a safety distance to avoid collisions, and the corresponding constraints can be expressed as follows: 
	\vspace{-3pt}
	\begin{equation}
		\mathcal{C}_1 : \left\|\mathbf{q}_u-\mathbf{q}_i\right\| \geq d_{\min }, i \neq u, \forall i \in \mathcal{U}, \forall u \in \mathcal{U},
		\vspace{-3pt}
	\end{equation}
	\noindent where $d_{\min}$ represents the minimum safety distance between UAV-SIMs. Likewise, the distance from the $u$-th UAV-SIM to the $m$-th user is given by $d_{m, u}=\left\|\mathbf{q}_u-\mathbf{p}_m\right\|$.
	
	\par Similar to \cite{DBLP:journals/twc/WuZZ18, DBLP:journals/corr/abs-2409-12870}, we consider that a UAV-SIM can only serve one user and a user can only be served by one UAV-SIM. Mathematically, we define a binary variable $\alpha_{m, u}$, where $\alpha_{m, u}=1$ means that there is an association between the $u$-th UAV-SIM and $m$-th user, i.e., $m$-th user can transmit data to the $u$-th UAV-SIM; otherwise, $\alpha_{m, u}=0$. The corresponding constraints can be written as follows:
	\vspace{-3pt}
	\begin{equation}
		\mathcal{C}_2: \sum\nolimits_{u=1}^U \alpha_{m, u} \leq 1, \forall m\in \mathcal{M},
	\end{equation}
	\begin{equation}
		\mathcal{C}_3: \sum\nolimits_{m=1}^M  \alpha_{m, u} \leq 1, \forall u\in \mathcal{U},
	\end{equation}
	\begin{equation}
		\mathcal{C}_4: \alpha_{m, u} \in \{0,1\}, \forall m\in \mathcal{M}, \forall u\in \mathcal{U}.
		\vspace{-3pt}
	\end{equation}
	
	\subsection{SIM Transmission Model}
	
	\par Let $\mathcal{L}=\{1, 2,..., L\}$ represent the set of metasurface layers, and $\mathcal{K}=\{1, 2,..., K\}$ represent the set of meta-atoms. Assume $\phi_k^l=\mathrm{e}^{j \theta_k^l}$ denotes the transmission coefficient imposed by the $k$-th meta-atom on the $l$-th metasurface layer, where $\theta_k^l$ denotes the corresponding phase shift, which satisfies $\theta_k^l \in[0,2 \pi), \forall k \in \mathcal{K}, \forall l \in \mathcal{L}$. Consequently, the phase shift matrix of the $l$-th metasurface layer is given as follows:
	\vspace{-3pt}
	\begin{equation}	\mathbf{\Phi}^l=\operatorname{diag}\left(\mathrm{e}^{j \theta_1^l}, \mathrm{e}^{j \theta_2^l}, \cdots, \mathrm{e}^{j \theta_K^l}\right) \in \mathbb{C}^{K \times K},\forall l \in \mathcal{L}.
		\vspace{-3pt}
	\end{equation}
	
	\par Without loss of generality, we assume that each metasurface is modeled as a uniformly planar array, and an isomorphic lattice structure is used to arrange all metasurface layers. As shown in Fig. \ref{System_model}(b), the distance between the $k$-th and $\tilde{k}$-th meta-atoms on the same metasurface layer can be written as follows \cite{DBLP:journals/jsac/AnXNAHYH23}:
	\vspace{-3pt}
	\begin{equation}
		s_{k, \tilde{k}}  =s_{e, k} \sqrt{\left(k_x-\tilde{k}_x\right)^2+\left(k_y-\tilde{k}_y\right)^2}, 
		\vspace{-3pt}
	\end{equation}
	\noindent where $s_{e, k}$ denotes the spacing between adjacent meta-atoms within the same metasurface layer. In addition, the indices of the $k$-th meta-atom in the $x$ and $y$ directions can be denoted as $k_x$ and $k_y$, respectively, which are defined as follows \cite{DBLP:journals/jsac/AnXNAHYH23}:
	\vspace{-3pt} 
	\begin{equation}
		k_x=\bmod \left(k-1, k_{\max }\right)+1, k_y=\left\lceil k / k_{\max }\right\rceil,
		\vspace{-3pt}
	\end{equation}
	\noindent where $k_{\max}$ is the number of meta-atoms on each row of the \pagebreak \clearpage
	\noindent metasurface. In this work, each metasurface is arranged in a square layout, and hence we have $K=k^2_{\max}$.
	
	\par Let $r$ represent the distance between adjacent metasurface layers in the SIM, and $T_{\mathrm{SIM}}$ is the thickness of the SIM. Accordingly, we have $r=T_{\mathrm{SIM}}/L$, and the distance between any two meta-atoms located in different adjacent metasurface layers can be represented as follows\cite{DBLP:journals/jsac/AnXNAHYH23}:
	\vspace{-4pt}
	\begin{equation}
		s_{k, \tilde{k}}^l=\sqrt{s_{k, \tilde{k}}^2+r^2}, \forall l \in \mathcal{L} /\{1\}.
		\vspace{-4pt}
	\end{equation}
	
	\par Denote $\mathbf{W}^l \in \mathbb{C}^{K \times K},\forall l \in  \mathcal{L} /\{1\}$ as the channel response from the $(l-1)$-th metasurface layer to the $l$-th metasurface layer, and $\mathbf{w}^{1\mathrm{H}}_u \in \mathbb{C}^{1 \times K}$ is the channel response from the output metasurface layer of the SIM to the $u$-th UAV. According to the Rayleigh-Sommerfeld diffraction theory\cite{liu2022programmable}, the response from the $\tilde{k}$-th meta-atom on the $(l-1)$-th metasurface layer to the $k$-th meta-atom on the $l$-th metasurface layer can be written as follows \cite{DBLP:journals/jsac/AnXNAHYH23}:
	\vspace{-4pt}
	\begin{equation}
		\label{W}
		w_{k, \tilde{k}}^l=\frac{A_t \cos \chi_{k, \tilde{k}}^l}{s_{k, \tilde{k}}^l}\left(\frac{1}{2 \pi s_{k, \tilde{k}}^l}-j \frac{1}{\lambda}\right) \mathrm{e}^{j 2 \pi s_{k, \tilde{k}}^l / \lambda}, \forall l \in \mathcal{L}/\{1\},
		\vspace{-4pt}
	\end{equation}
	\noindent where $s_{k, \tilde{k}}^l$ is the transmission distance, $A_t$ denotes the area of each meta-atom and $\chi_{k, \tilde{k}}^l$ is the angle between the propagation direction and the normal direction of the $(l-1)$-th metasurface layer. Similarly, the $k$-th entry $w^1_{k,u}$ of $\mathbf{w}^{1}_u$ can be obtained using Eq. (\ref{W}).
	
	\par Therefore, the equivalent channel response at the $u$-th UAV-SIM can be written as follows \cite{DBLP:journals/twc/PapazafeiropoulosAKRC24}:
	\vspace{-4pt}
	\begin{equation}
			\mathbf{G}_{u}=\boldsymbol{\Phi}^L_{u} \mathbf{W}^L_{u} \boldsymbol{\Phi}^{L-1}_{u} \mathbf{W}^{L-1}_{u} \ldots \boldsymbol{\Phi}^2_{u} \mathbf{W}^2_{u} \boldsymbol{\Phi}^1_{u}  \in \mathbb{C}^{K \times K} .
			\vspace{-4pt}
	\end{equation}
	\addtolength{\topmargin}{0.01in}
	
	\par Let $\mathbf{h}_{m,u} \in \mathbb{C}^{K \times 1}$ denote the baseband equivalent channel from the $m$-th user to the input metasurface layer of the $u$-th UAV-SIM. Without loss of generality, a quasi-static flat-fading model is adopted for all channels, and hence $\mathbf{h}_{m,u}  \sim \mathcal{C N}\left(\mathbf{0}, \rho_0 d_{m, u}^{-2} \mathbf{R}\right), \forall m\in\mathcal{M}, \forall u\in\mathcal{U}\label{h}$ according to the correlated Rayleigh fading distribution, where $\rho_0$ denotes the channel power at the reference distance of $1$ m, and $\mathbf{R} \in \mathbb{C}^{K \times K}$ represents the spatial correlation matrix of the SIM. In an isotropic scattering environment, the spatial correlation matrix of the SIM for the far-field propagation is $\mathbf{R}_{k, \tilde{k}}=\operatorname{sinc}\left(2 s_{k, \tilde{k}} / \lambda\right), \forall \tilde{k} \in \mathcal{K},\forall k \in \mathcal{K},$ where $s_{k, \tilde{k}}$ represents the corresponding meta-atom spacing.
	
	\par  Then, the composite signal received by the $u$-th UAV-SIM can be written as follows\cite{DBLP:journals/wcl/YaoAGRY24}:
	\vspace{-4pt} 
	{\begin{equation}
			r_{m,u} = \mathbf{w}_{u}^{1\mathrm{H}} \mathbf{G}_{u}^\mathrm{H} \sum\nolimits_{m=1}^M \mathbf{h}_{m,u}\sqrt{p_{m}} g_{m}+n_u,\forall m\in \mathcal{M}, \forall u\in \mathcal{U},
			\vspace{-4pt} 
	\end{equation}
	\noindent where $p_m$ is the transmit power of the $m$-th user, $g_{m}$ is the signal from the $m$-th user and $n_u  \sim \mathcal{C N}\left(0, \sigma_u^2\right)$ denotes the additive white Gaussian noise (AWGN), as the multiuser interference is automatically mitigated as EM waves pass through the well-designed SIMs. Therefore, the signal-to-interference-plus-noise ratio (SINR) at the $u$-th UAV-SIM from the $m$-th user is given by \cite{DBLP:conf/icc/AnRDY23}:
	\begin{equation}	   \gamma_{m,u}=\frac{\left|\mathbf{w}_u^{1\mathrm{H}} \mathbf{G}_{u}^\mathrm{H} \mathbf{h}_{m,u}\right|^2 p_m }{\sum_{m^{\prime} \neq m}^M\left|\mathbf{w}_{u}^{1\mathrm{H}} \mathbf{G}_{u}^\mathrm{H} \mathbf{h}_{m^{\prime},u}\right|^2 p_{m^{\prime}}+\sigma_u^2},\forall m\in \mathcal{M}, \forall u\in \mathcal{U},
	\end{equation}
	\noindent where $\sigma_u^2$ represents the average noise power of the $u$-th UAV. Then, the available rate at the $u$-th UAV-SIM from the $m$-th user is calculated as follows:
	\vspace{-0.5em} 
	\begin{equation}
		R_{m,u} =  \log_2(1+\gamma_{m,u}).
	\end{equation}
	
	\par As a result, the network capacity can be expressed as:
	\begin{equation}	R=\sum\nolimits_{m=1}^M \sum\nolimits_{u=1}^U\alpha_{m, u}R_{m,u}.
	\end{equation}
	
	\subsection{Problem Formulation}
	
	\par The optimization objective is to maximize the network capacity by jointly adjusting the association between UAV-SIMs and users, the locations of UAV-SIMs, and the phase shifts of UAV-SIMs. By defining $\mathbf{A}\triangleq\left\{\alpha_{m,u}, \forall m\in \mathcal{M}, \forall u\in \mathcal{U}\right\}$, $\mathbf{Q}\triangleq \left\{\mathbf{q}_u,  \forall u\in \mathcal{U} \right\}$, $\boldsymbol{\theta}^l \triangleq\left[\theta_1^l, \theta_2^l, \cdots, \theta_K^l\right]^{\mathrm{T}}$ and $\boldsymbol{\vartheta} \triangleq\left\{\boldsymbol{\theta}^1, \boldsymbol{\theta}^2, \cdots, \boldsymbol{\theta}^L\right\}$, the USBJOP can be formulated as follows:
	\begin{subequations}\label{eq:2}
		\begin{align}
			\text {(USBJOP) }:\underset{\mathrm{\mathbf{A},\mathbf{Q},\boldsymbol{\vartheta}}}{\max}\quad  &R  \\
			\text { s.t. } \quad
			& \mathcal{C}_1-\mathcal{C}_4,\\ 
			& \mathcal{C}_5:\theta_k^l \in[0,2 \pi), \forall k \in \mathcal{K}, \forall l \in \mathcal{L}. \label{phase shift constraint1}
		\end{align}
	\end{subequations}
	\noindent where $\mathcal{C}_5$ is the phase shift constraint. As can be seen, USBJOP contains both the continuous variables and the discrete variables, and hence it is a non-linear mixed integer optimization problem, which is non-convex and NP-hard\cite{BURER201297}. 
	\vspace{-0.5em} 
	
	%
	%
	\section{Proposed Solution}
	\label{section3}
	
	\par Due to the intractability of the formulated USBJOP, we decompose it into three sub-optimization problems, which are AUUOP, ULOP, and USPSOP. Then, an AO strategy is adopted to solve these three sub-optimization problems, which is shown in Algorithm \ref{alg:2}. Finally, the computational complexity of the AO strategy is analyzed. The details for solving each sub-optimization problem are as follows.
	\addtolength{\topmargin}{0.04in}
	\begin{algorithm}
		\SetAlgoLined
		\SetAlgoNlRelativeSize{-1} 
		Let $\tau=0$ and initialize $\boldsymbol{\vartheta}$ and $\mathbf{Q}$;\\
		\Repeat{$R^{\tau}-R^{\tau-1}\leq\epsilon$ or $\tau=\tau_{\max}$}{
			Update $\mathbf{A}$ to solve AUUOP using CVX tool;\\
			Update $\mathbf{Q}$ to solve ULOP using CVX tool;\\
			Update $\boldsymbol{\vartheta}$ to solve USPSOP using Algorithm \ref{algo:3};\\
			$\tau=\tau+1$;\\
		}
		\caption{AO strategy for solving USBJOP\label{alg:2}}
	\end{algorithm}
	
	\subsection{AUUOP} 
	
	\subsubsection{AUUOP Formulation}
	\label{alpha optimization}
	
	\par Given the UAV-SIM phase shifts and the locations of the UAV-SIMs $\{\boldsymbol{\vartheta},\mathbf{Q}\}$, the sub-optimization problem AUUOP is shown as follows: 
	\begin{subequations}
		\begin{align}
			\text {(AUUOP)}:\underset{\mathrm{\mathbf{A}}}{\max}\quad & R\\
			\text { s.t. } \quad & \mathcal{C}_2-\mathcal{C}_4.
		\end{align}
	\end{subequations}
	
	\par To make AUUOP more tractable, we relax $\mathcal{C}_4$, and then $\alpha_{m, u}$ is transformed into a continuous variable, which satisfies $0 \leq \alpha_{m,u} \leq 1, \forall m\in \mathcal{M}, \forall u\in \mathcal{U}$. Hence, the modified AUUOP (M-AUUOP) can be rewritten as follows:
	\begin{subequations}
		\begin{align}
			\underset{\mathrm{\mathbf{A}}}{\max}\quad & R\\
			\text { s.t. } \quad
			& 0 \leq \alpha_{m,u} \leq 1, \forall m\in \mathcal{M}, \forall u\in \mathcal{U},\\
			& \mathcal{C}_2,\mathcal{C}_3.
		\end{align}
	\end{subequations}
	
	\subsubsection{AUUOP Solution}
	
	\par It is evident that M-AUUOP is a standard linear programming problem, which can be effectively solved by the CVX tool \cite{grant2014cvx}. 
	
	\par \textbf{\textit{Remark 1:}} We reconstruct the obtained solution after solving AUUOP. Specifically, for the $u$-th UAV-SIM, we associate the user with the maximum rate and dissociate all other users. At this point, the association variable $\alpha$ is reinstated as a tight binary constraint. This ensures the feasibility of the solutions obtained for the formulated sub-optimization problem.
	\vspace{-6pt}
	
	\subsection{ULOP}
	
	\subsubsection{ULOP Formulation}
	
	\par Given the UAV-SIM phase shifts and the  association between UAV-SIMs and users $\{\boldsymbol{\vartheta},\mathbf{A}\}$, ULOP can be written as follows:
	\begin{subequations}
		\label{al2}
		\begin{align}    \text{(ULOP)}:\underset{\mathrm{\mathbf{Q}}}{\max}\quad & R\\
			\text { s.t. } \quad
			& \mathcal{C}_1. \label{20b}
		\end{align}
	\end{subequations}
	
	\subsubsection{ULOP Solution}
	
	\par Due to the non-convexity of $\mathcal{C}_1$, ULOP is still non-convex. Hence, we adopted the successive convex approximation (SCA) approach to deal with the constraint \cite{DBLP:journals/twc/WuZZ18}. By defining $\mathbf{\tilde h}_{m,u}\sim \mathcal{C} \mathcal{N}\left(\mathbf{0},\mathbf{R}\right),\forall m\in \mathcal{M}, \forall u\in \mathcal{U}$, the relevant channel $\mathbf{h}_{m,u}$ can be rewritten as follows:
	\begin{equation}
		\mathbf{h}_{m, u}=\sqrt{\beta_{m, u}} \tilde{\mathbf{h}}_{m, u}.
	\end{equation}
	Hence, the sub-optimization problem ULOP can be rewritten as follows:
	\begin{equation}
			\begin{split}
				\label{R}
				R= & \sum\nolimits_{m=1}^M \sum\nolimits_{u=1}^U  \alpha_{m,u} \Bigg(\hat{R}_{m,u}-\\
				&\log _2\bigg(\sum\nolimits_{m^{\prime} \neq m}^M \frac{ \rho_0 p_{m^{\prime}}}{\left\|\mathbf{q}_u-\mathbf{p}_{m^{\prime}}\right\|^2}|\mathbf{w}_u^{1\mathrm{H}} \mathbf{G}_{u}^{\mathrm{H}} \mathbf{\tilde h}_{m^{\prime},u}|^2+\sigma_u^2\bigg) \Bigg),
			\end{split}
	\end{equation}
	where $\hat{R}_{m,u} = \log_2\left( \sum\nolimits_{l=1}^M \frac{ \rho_0 p_l \left| \mathbf{w}_u^{1\mathrm{H}} \mathbf{G}_{u}^{\mathrm{H}} \mathbf{\tilde h}_{l,u} \right|^2 }{ \left\| \mathbf{q}_u - \mathbf{p}_l \right\|^2 } + \sigma_u^2 \right)$.
	
	\par Considering Eq. (\ref{R}) is still non-convex with respect to $\mathbf{q}_u$, we let $\mathbf{S}=\{S_{m^{\prime},u}=\left\|\mathbf{q}_u-\mathbf{p}_{m^{\prime}}\right\|^2,  m^{\prime}\neq m,\forall m^{\prime} \in \mathcal{M},\forall u \in \mathcal{U}\}$ to relax the problem. Therefore, ULOP can be reformulated as follows:
	\begin{subequations}\label{21}
		\begin{align}		\underset{\mathrm{\mathbf{Q},\mathbf{S}}}{\max}\quad& R= \sum\nolimits_{m=1}^M \sum\nolimits_{u=1}^U  \alpha_{m,u}\Bigg(\hat{R}_{m,u}-\\
			&\phantom{R=}\log _2\bigg(\sum\nolimits_{m^{\prime} \neq m}^M \frac{\rho_0 p_{m^{\prime}}}{S_{m^{\prime},u}}|\mathbf{w}_u^{1\mathrm{H}} \mathbf{G}_{u}^{\mathrm{H}} \mathbf{\tilde h}_{m^{\prime},u}|^2+\sigma_u^2\bigg)\Bigg),\notag\\
			\text { s.t. } \quad
			& \mathcal{C}_1 \label{24c},\\
			& \mathcal{C}_6:0\leq S_{m^{\prime},u}\leq\left\|\mathbf{q}_u-\mathbf{p}_{m^{\prime}}\right\|^2.\label{24b}
		\end{align}
	\end{subequations}
	
	\par Given a local point $\mathbf{Q}^\tau= \left\{\mathbf{q}_u^\tau, \forall u\in \mathcal{U}, \right\}$ at the $\tau$-th iteration, the lower bound for $\hat{R}_{m,u}$ is readily derived by the first-order Taylor expansion as follows:
	\begin{equation}
		\sum_{m=1}^M-A_{m,u}^\tau\left(\left\|\mathbf{q}_u-
		\mathbf{p}_m\right\|^2-\left\|\mathbf{q}_u^\tau-\mathbf{p}_m\right\|^2\right)+B_{m,u}^\tau \triangleq \hat{R}^{\prime}_{m,u},
	\end{equation}
	where $A_{m,u}^\tau$ and $B_{m,u}^\tau$ can be written as		
	\begin{align}
		&A_{m,u}^\tau=\frac{\frac{\rho_0 p_m}{\left(\left\|\mathbf{q}_u^\tau-\mathbf{p}_m\right\|^2\right)^2}\left|\mathbf{w}_u^{1\mathrm{H}} \mathbf{G}_{u}^{\mathrm{H}} \mathbf{\tilde h}_{m,u}\right|^2 \log _2(\mathrm{e})}{\sum_{l=1}^M \frac{ \rho_0 p_l}{\left\|\mathbf{q}_u^\tau-\mathbf{p}_l\right\|^2}\left|\mathbf{w}_u^{1\mathrm{H}} \mathbf{G}_{u}^{\mathrm{H}} \mathbf{\tilde h}_{l,u}\right|^2+\sigma_{u}^2},\\
		&B_{m,u}^\tau=\log _2\left(\sum\nolimits_{l=1}^M \frac{\rho_0 p_l}{\left\|\mathbf{q}_u^\tau-\mathbf{p}_l\right\|^2}\left|\mathbf{w}_u^{1\mathrm{H}} \mathbf{G}_{u}^{\mathrm{H}} \mathbf{\tilde h}_{l,u}\right|^2+\sigma_{u}^2\right).
	\end{align}
	
	\par Similarly, we can obtain a tighter constraint for $\mathcal{C}_1$ and $\mathcal{C}_6$ at the given $\mathbf{Q^\tau}$ as
	\begin{align}
		\mathcal{C}&_7:0\leq S_{m^{\prime},u} \leq\left\|\mathbf{q}_u^\tau-\mathbf{p}_{m^{\prime}}\right\|^2+2\left(\mathbf{q}_u^\tau-\mathbf{p}_{m^{\prime}}\right)^{\mathrm{T}}\left(\mathbf{q}_u-\mathbf{q}_u^\tau\right),\\
		&\mathcal{C}_8:d_{\min }^2 \leq-\left\|\mathbf{q}_u^\tau-\mathbf{q}_i^\tau\right\|^2 +2\left(\mathbf{q}_u^\tau-\mathbf{q}_i^\tau\right)^{\mathrm{T}}\left(\mathbf{q}_u-\mathbf{q}_i\right).
	\end{align}
	
	\par Therefore, the modified ULOP (M-ULOP) is approximately formulated as the following problem:
	\begin{subequations}\label{27}
		\begin{align}
			\underset{\mathrm{\mathbf{Q},\mathbf{S}}}{\max}\quad &R= \sum\nolimits_{m=1}^M \sum\nolimits_{u=1}^U  \alpha_{m,u}\Bigg(\hat{R}^{\prime}_{m,u}- \\ 
			&\phantom{R=}\log _2\bigg(\sum\nolimits_{m^{\prime} \neq m}^M \frac{\rho_0 p_{m^{\prime}}}{S_{m^{\prime},u}}|\mathbf{w}_u^{1\mathrm{H}} \mathbf{G}_{u}^{\mathrm{H}} \mathbf{\tilde h}_{m^{\prime},u}|^2+\sigma_u^2\bigg)\Bigg)\notag\\
			\text { s.t. } \quad
			& \mathcal{C}_7,\mathcal{C}_8.
		\end{align}
	\end{subequations}
	
	\par Since M-ULOP is jointly concave with respect to $\mathbf{q}_u$ and $S_{m^{\prime},u}$, it is a convex function in this work. Additionally, as $\mathcal{C}_7$ is linear and $\mathcal{C}_8$ is a convex quadratic constraint, M-ULOP is a convex problem that can also be addressed by the CVX tool. Apparently, the solution obtained from M-ULOP is feasible for the ULOP. Thus, we approximate the optimal solution of M-ULOP as the solution to ULOP.
	
	\subsection{USPSOP}\label{phase optimization algorithm}
	
	\subsubsection{USPSOP Formulation}
	Given the association between UAV-SIMs and users and the locations of the UAV-SIMs $\{\mathbf{A},\mathbf{Q}\}$, the sub-optimization problem USPSOP can be written as follows:
	\vspace{-3pt}
	\begin{subequations}
		\begin{align}
			\text { (USPSOP) }:\underset{\mathrm{\boldsymbol{\vartheta}}}{\max}\quad & R\\
			\text { s.t. } \quad
			& \mathcal{C}_5. 
		\end{align}
		\vspace{-8pt}
	\end{subequations}
	
	\subsubsection{USPSOP Solution}
	
	\par Since USPSOP is also non-convex, it remains challenging to solve. To tackle this issue, we employ a layer-by-layer iterative phase shift optimization method to obtain the suboptimal solution \cite{li2024stacked}.
	
	\par For the $u$-th UAV-SIM, we fix the phase shifts of the other $(L-1)$ layers to optimize the phase shift $\boldsymbol{\Phi}^l_{u}$ of the $l$-th layer. The channel gain $\mathbf{w}_u^{1\mathrm{H}} \mathbf{G}_{u}^{\mathrm{H}} \mathbf{h}_{m,u}$ can be expressed as  $\mathbf{F}^{l\mathrm{H}}_u\boldsymbol{\Phi}^{l\mathrm{H}}_{u}\mathbf{L}^{l\mathrm{H}}_u$ in conjunction with
	\vspace{-3pt}
	\begin{equation}\label{29}
			\mathbf{F}^{l\mathrm{H}}_u=\mathbf{w}_u^{1\mathrm{H}}\boldsymbol{\Phi}^{1\mathrm{H}}_{u}\mathbf{W}^{2\mathrm{H}}_{u}\boldsymbol{\Phi}^{2\mathrm{H}}_{u}\ldots\mathbf{W}^{l\mathrm{H}}_{u},
		\vspace{-3pt}
	\end{equation}
	and 
	\vspace{-3pt}
	\begin{equation}\label{30}
			\mathbf{L}^{l\mathrm{H}}_u=\mathbf{W}_u^{(l+1)\mathrm{H}} \boldsymbol{\Phi}_u^{(l+1)\mathrm{H}} \ldots \mathbf{W}_u^{L\mathrm{H}} \boldsymbol{\Phi}_u^{L\mathrm{H}} \mathbf{h}_{m,u}.
		\vspace{-3pt}
	\end{equation} 
	Therefore, $\boldsymbol{\Phi}^l_{u}$ can be optimized as follows:
	\vspace{-3pt}
	\begin{equation}\label{31}
			\boldsymbol{\Phi}_u^{l}=\operatorname{diag}\left\{\mathrm{e}^{-j\left(\angle \mathbf{F}^{{l}}_u-\angle \mathbf{L}^{l\mathrm{H}}_u\right)}\right\}.
		\vspace{-3pt}
	\end{equation}
	\addtolength{\topmargin}{0.01in}
	
	\par Accordingly, $\boldsymbol{\Phi}_u^1$,  $\boldsymbol{\Phi}_u^2$, ...,  $\boldsymbol{\Phi}_u^L$ can be sequentially optimized using this approach. We will perform $\kappa_{\max}$ iterations to ensure that the phase shift optimization achieves optimal results. The details of the layer-by-layer iterative phase shift optimization method are shown in Algorithm \ref{algo:3}. 
	\vspace{-10pt} 
	\begin{algorithm}
		\DontPrintSemicolon
		\SetAlgoNlRelativeSize{-1} 
		Let $\kappa=1$ and initialize  $\boldsymbol{\Phi}_u^1$,  $\boldsymbol{\Phi}_u^2$, ...,  $\boldsymbol{\Phi}_u^L$ randomly;\\
		\For{$u=1$ to $U$}{
			\For{$\kappa=1$ to $\kappa_{\max}$}{
				\For{$l=1$ to $L$}{
					Calculate $\mathbf{F}^{l\mathrm{H}}_u$ according to Eq. (\ref{29});\\	
					Calculate $\mathbf{L}^{l\mathrm{H}}_u$ according to Eq. (\ref{30});\\
					Calculate $	\boldsymbol{\Phi}_u^{l}$ according to Eq. (\ref{31});\\
				}
			}
		}
		\caption{The layer-by-layer iterative phase shift optimization method \label{algo:3}}
	\end{algorithm}
	\vspace{-18pt}
	
	\subsection{Computational Complexity Analysis}
	
	\par For AUUOP, $\mathrm{\mathbf{A}}$ has $MU$ variables. Therefore, according to the analysis in \cite{DBLP:journals/tmc/WangWPXAN22}, the required number of iterations is $\sqrt{MU} \log _2\big({1}/{\varepsilon_1}\big)$, where $\varepsilon_1$ represents the required search accuracy for solving AUUOP with CVX. In addition, the number of constraints in AUUOP is $MU+M+U$. Thus, the computational complexity for AUUOP is calculated as $\mathcal{O}\Big((MU)^{3.5}\log _2\big({1}/{\varepsilon_1}\big)\Big)$. For ULOP, the total number of variables for $\mathrm{\mathbf{Q}}$ and $\mathrm{\mathbf{S}}$ is $(M+2)U$. Similarly, the required number of iterations is $\sqrt{(M+2)U} \log _2\big({1}/{\varepsilon_2}\big)$, where $\varepsilon_2$ represents the required search accuracy for solving ULOP with CVX. The number of constraints in ULOP is ${U(U-1)}/{2}+(M-1)U$. Thus, the computational complexity for ULOP is calculated as $\mathcal{O}\Big((MU)^{2.5}(U^2+MU)\log _2\big({1}/{\varepsilon_2}\big)\Big)$. For USPSOP, the calculation of $\mathbf{F}^{l\mathrm{H}}_u$ in line 5 of Algorithm \ref{algo:3} requires $2(l-1)K^2$ floating-point multiplications, while calculating $\mathbf{L}^{l\mathrm{H}}_u$ in line 6 requires $2(L-l)K^2$ floating-point multiplications. Therefore, the computational complexity of Algorithm \ref{algo:3} is $\mathcal{O}(2K^2UL(L-1)\kappa_{\max})$. In summary, the computational complexity of Algorithm \ref{alg:2} is approximately $\mathcal{O}\bigg(\Big(2K^2UL(L-1)\kappa_{\max}+(MU)^{3.5}(\log _2\big({1}/{(\varepsilon_1\varepsilon_2)}\big)\big)\Big)\tau_{\max}\bigg)$ in the worst case. It is proven that the proposed AO strategy is viable and can efficiently address USBJOP.
	\vspace{-5pt}
	
	%
	%
	\section{Simulation Results}\label{sec4}
	\vspace{-2pt}
	
	\subsection{Simulation Setup}
	
	\par In this section, numerical results are presented to demonstrate the effectiveness of the proposed AO strategy. For parameter setup, $M=5$ ground users are distributed randomly in a square area with the size of $1000$ m $\times 1000$ m. Moreover, $3$ UAV-SIMs are flying in an aerial region at a fixed height $H=50$ m. The safety distance is set as $d_{\mathrm{min}}=100$ m, and the thickness of the SIM is set to $T_{\mathrm{SIM}}=5\lambda$. Therefore, for an $L$-layer SIM, the spacing between layers is $r=T_{\mathrm{SIM}}/L$. Additionally, we specify that the meta-atoms of the SIM are arranged in a square configuration, with each meta-atom having a size of $A_t=(\lambda/2)^2$. In this paper, we consider the radio frequency of $28$ GHz and the corresponding wavelength is $\lambda=10.7$ mm. Furthermore, the number of meta-atoms for each layer of the SIM is set to $K=36$. The noise power is assumed to be $\sigma^2_u=-110\text{ dBm}, \forall u\in \mathcal{U}$, and the transmit power of each user is set to $500\text{ mW}$. Moreover, the channel gain at a reference distance of 1 m is set to $\rho_0=(\lambda/4\pi)^2$. We set the threshold $\epsilon$ for the AO strategy as $10^{-6}$, and the maximum iteration of AO is set as $\tau_{\max}=50$. In addition, for Algorithm \ref{algo:3}, we set the number of iterations as $\kappa_{\max}=10$.
	\vspace{-5pt}
	
	\subsection{Benchmarks}
	\vspace{-2pt}
	
	\par For comparison, we employ several benchmarks, which are detailed as follows:
	\begin{itemize}
		\item \textbf{Without SIM}:
		We consider the scenario where each UAV is not equipped with an SIM, serving uplink communications for users.
		\item \textbf{Random deployment (RD)}:
		In RD, the solution of USBJOP is generated randomly. To ensure RD performance, we generate $100$ solutions and select the best one.
		\item \textbf{Uniform deployment (UD)}:
		We divide the entire horizontal area into $U$ sub-regions, with each UAV-SIM located at the center of its corresponding sub-region, while maintaining a fixed height $H$. For the remaining two sub-optimization problems, we continue to utilize the corresponding solution in Section \ref{section3}. This strategy is commonly used in UAV deployment scenarios \cite{pan2023joint}.
		\item \textbf{Evolutionary algorithms}:
		We select two classic evolutionary algorithms, which are differential evolution (DE) and particle swarm optimization (PSO) \cite{pan2023joint}  for comparison \cite{pan2023joint}, and set the iteration number to $50$. Specifically, the USBJOP is still decomposed into three sub-optimization problems, and each sub-optimization problem is addressed by the corresponding evolutionary algorithms. 
	\end{itemize}
	\vspace{-6pt}
	
	\subsection{Optimization Results}
	
	\par In this section, we present a visualization of the optimization results, verify the convergence of the proposed AO strategy, and evaluate the performance.
	\begin{figure}[h]
		\centering
		\begin{minipage}[t]{\linewidth}
			\centering
			\includegraphics[width=0.9\linewidth]{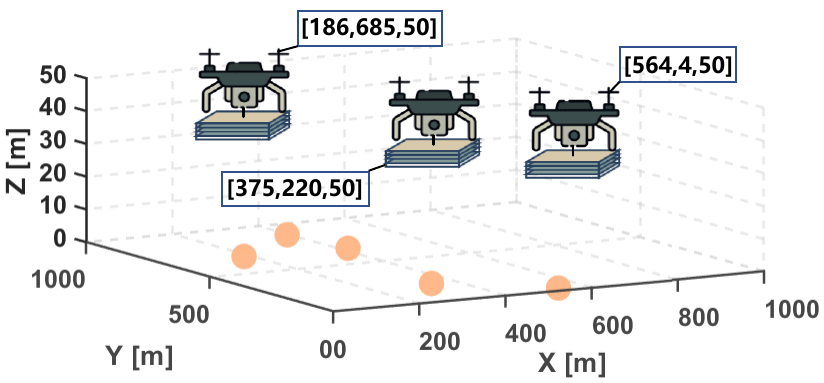}
			\caption{3D deployment results obtained by the proposed AO strategy for solving USBJOP.}
			\label{3D optimization results obtained by proposed AO strategy for solving USBJOP}
		\end{minipage}
		\vspace{-15pt}
	\end{figure}
	
	\par Fig. \ref{3D optimization results obtained by proposed AO strategy for solving USBJOP} shows the 3D deployment results obtained by the proposed AO strategy for solving USBJOP, where $L=3$. It is evident from the figure that the UAV-SIMs maintain the safety distance by using the proposed AO strategy. Moreover, they tend to gather in areas with high concentrations of users, which verifies the performance of the proposed AO strategy.
	
	\par Fig. \ref{R-Iteration} shows the convergence of the proposed AO strategy compared to PSO, DE, UD and UAV communications without SIM, where we consider two setups with $L=6$ and $L=7$, as the optimization results of AO gradually converge when $L = 7$. As can be seen, using UAV-SIM can obtain a greater network capacity compared to communications without SIM. Besides, for the proposed AO strategy, as the number of iterations increases, the network capacity steadily improves, indicating that the quality of the obtained solution is becoming better. Moreover, the proposed AO strategy obtains a faster convergence speed than PSO and DE, and provides better solutions compared to UD.
	
	\par Fig. \ref{R-L} shows the network capacity under different 
	number of metasurface layers. Obviously, the proposed AO strategy always obtains the maximum network capacity compared to other benchmarks as increasing the number of metasurface layers $L$ from $1$ to $8$. Moreover, the network capacity tends to converge when $L$ increases, achieving its maximum at approximately $L=7$, thereby indicating that the multi-user interference cancellation capability of the SIM is positively correlated with the number of layers $L$ within a certain range. It is worth mentioning that the proposed AO strategy outperforms the suboptimal UD, with more than double the network capacity when $L=7$.
	
	%
	%
	\section{Conclusions}\label{sec5}
	In this paper, the UAV-SIM-assisted uplink communication system has been investigated. First, we have considered deploying multiple UAV-SIMs to serve the ground users. Then, we have formulated the USBJOP to jointly adjust the association between UAV-SIMs and users, the locations of UAV-SIMs, and the phase shifts of UAV-SIMs to maximize the network capacity. Due to the complexity of the USBJOP, we have decomposed it into three sub-optimization problems, which are AUUOP, ULOP, and USPSOP, and then proposed the AO strategy to solve them. Simulation results have indicated that the proposed strategy outperforms some benchmarks, and the AO strategy obtains a better convergence performance.
	\begin{figure}[tb]
		\centering
		\begin{minipage}[t]{\linewidth}
			\centering
			\includegraphics[width=2.9in]{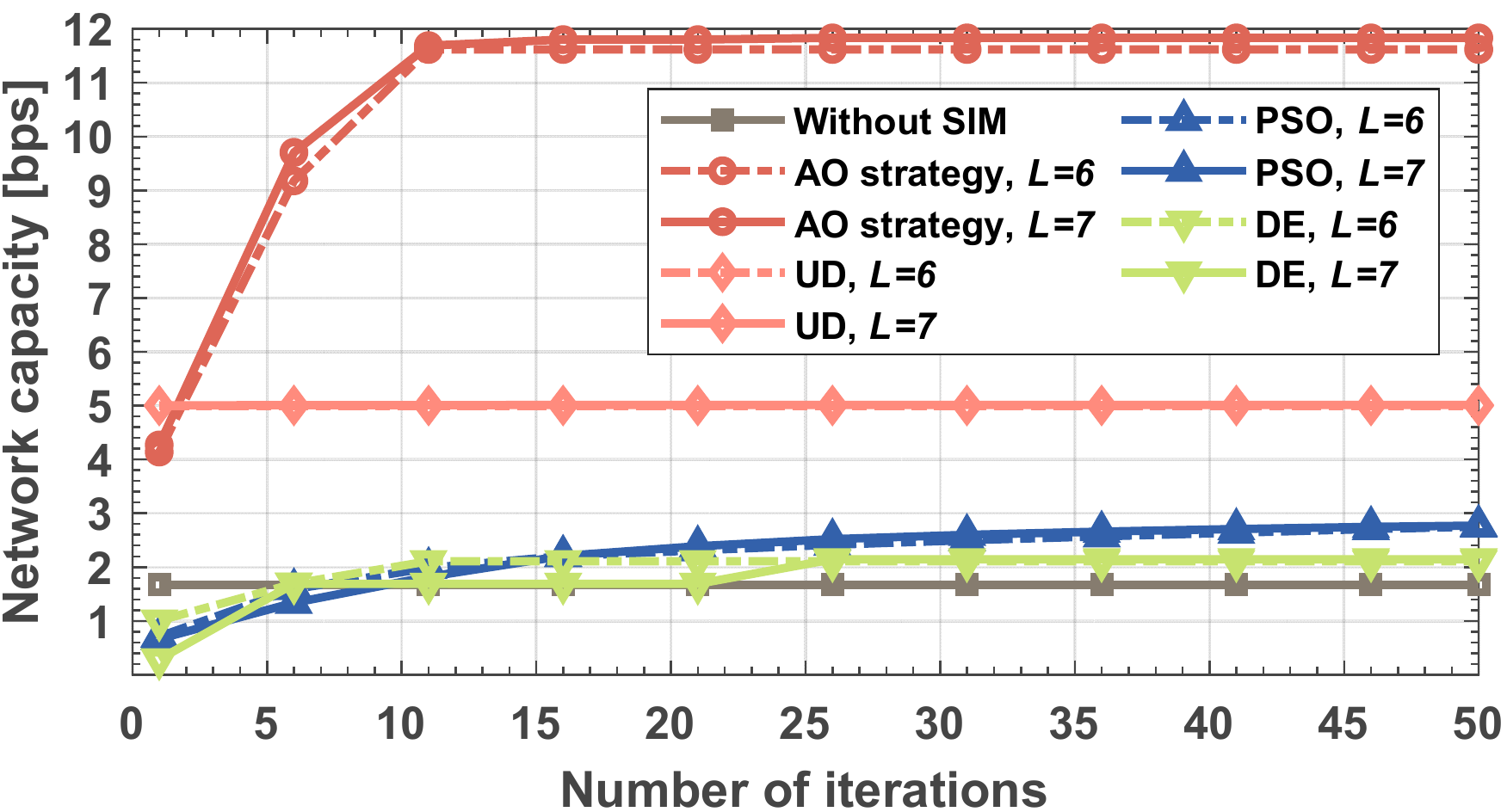}
			\caption{Performance comparison of different benchmarks.}
			\label{R-Iteration}
		\end{minipage}
		\begin{minipage}[t]{\linewidth}
			\centering
			\includegraphics[width=2.9in]{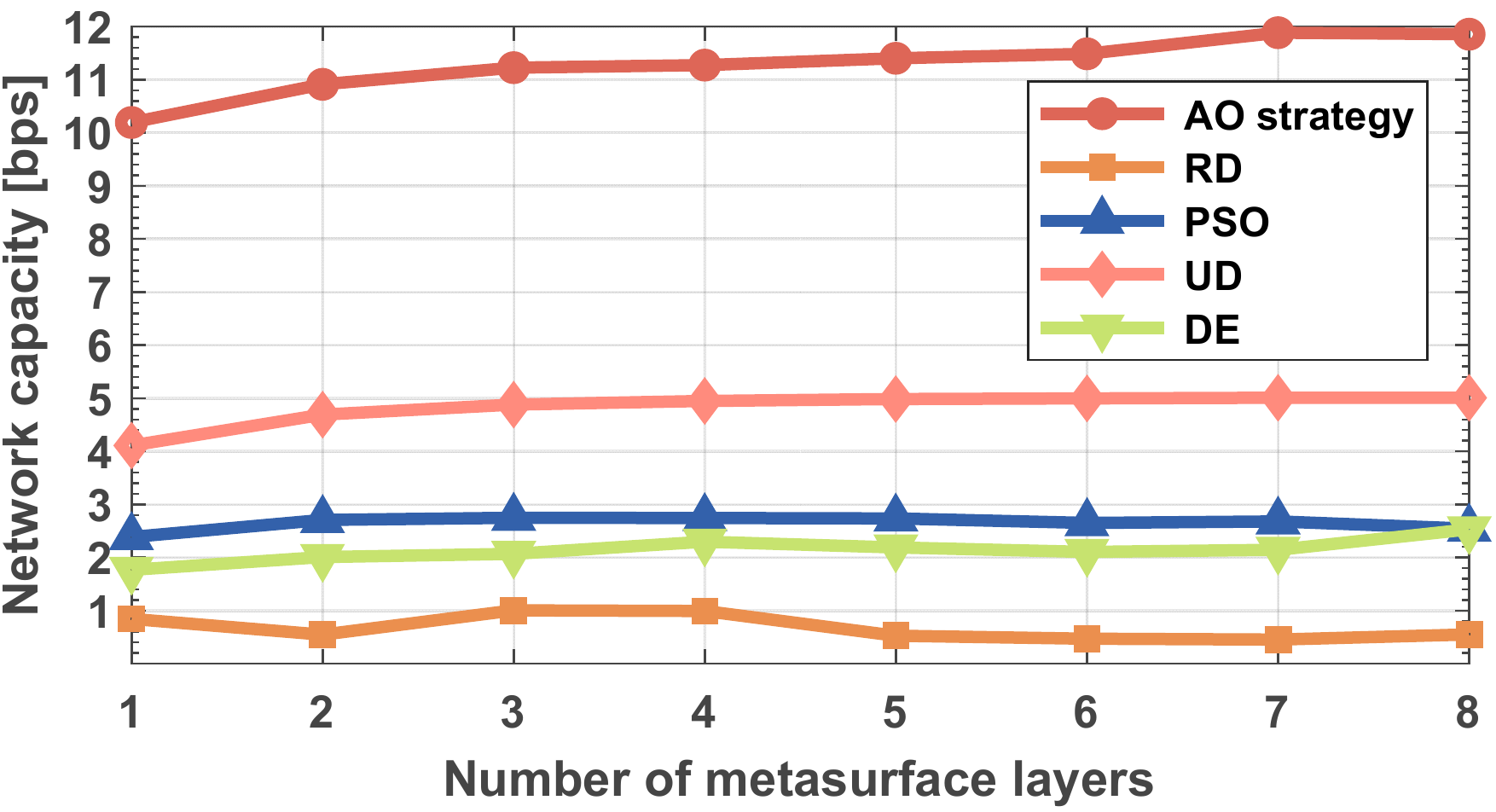} 
			\caption{Network capacity $R$ versus the number of metasurface layers $L$.}
			\label{R-L}
		\end{minipage}
		\vspace{-15pt}
	\end{figure}
	\bibliography{reference}
	\vfill
	\bibliographystyle{ieeetr}
\end{document}